\documentclass[12pt]{article}
\usepackage{graphicx,epsfig}
\usepackage{color}
\title{
Magnetization plateaus and phase diagrams of the extended Ising model 
on the Shastry-Sutherland lattice: Effects of long-range interactions}
\author{Pavol Farka\v sovsk\'y and Lubom\'ira Regeciov\'a\\
Institute  of  Experimental  Physics,  Slovak   Academy   of
Sciences\\
Watsonova 47, 040 01 Ko\v {s}ice, Slovakia}
\date{}
\begin{document}
\baselineskip=24pt
\maketitle

\begin{abstract}
Magnetization plateaus and phase diagrams of the extended Ising model on 
the Shastry-Sutherland lattice with the first $(J_1)$, second $(J_2)$, third 
$(J_3)$ fourth $(J_4)$ and fifth $(J_5)$ nearest-neighbour spin couplings 
are studied by the classical Monte Carlo method. It is shown that 
switching on $J_4$ and $J_5$ interactions (in addition to usually
considered $J_1, J_2$ and $J_3$ interactions) changes significantly
the picture of magnetization processes found for $J_4=J_5=0$ and 
leads to stabilization of new macroscopic magnetic phases (plateaus)
with fractional magnetization. In particular, it is found that
combined effects of $J_4$ and $J_5$ interactions generate the
following sequence of plateaus with the fractional magnetization: 
$m/m_s$=1/9, 1/6, 2/9, 1/3, 4/9, 1/2, 5/9 and 2/3. The results obtained 
are consistent with experimental measurements of magnetization curves 
in selected rare-earth tetraborides.
\end{abstract}

\newpage
\section{Introduction}
A spin system is frustrated when all local interactions between spin pairs
cannot be satisfied at the same time. Frustration can arise from competing
interactions or/and from a particular geometry of the lattice, as seen in
the triangular lattice. The Shastry-Sutherland lattice (SSL) was considered
more than 30 years ago by Shastry and Sutherland~\cite{Shastry} as an
interesting example of a frustrated quantum spin system with an exact ground
state. It can be described as a square lattice with antiferromagnetic 
couplings $J_1$ between nearest neighbours and additional 
antiferromagnetic couplings $J_2$ between next-nearest neighbours 
in every second square (see Fig.~1a). 
\begin{figure}[h!]
\begin{center}
\includegraphics[width=7.2cm]{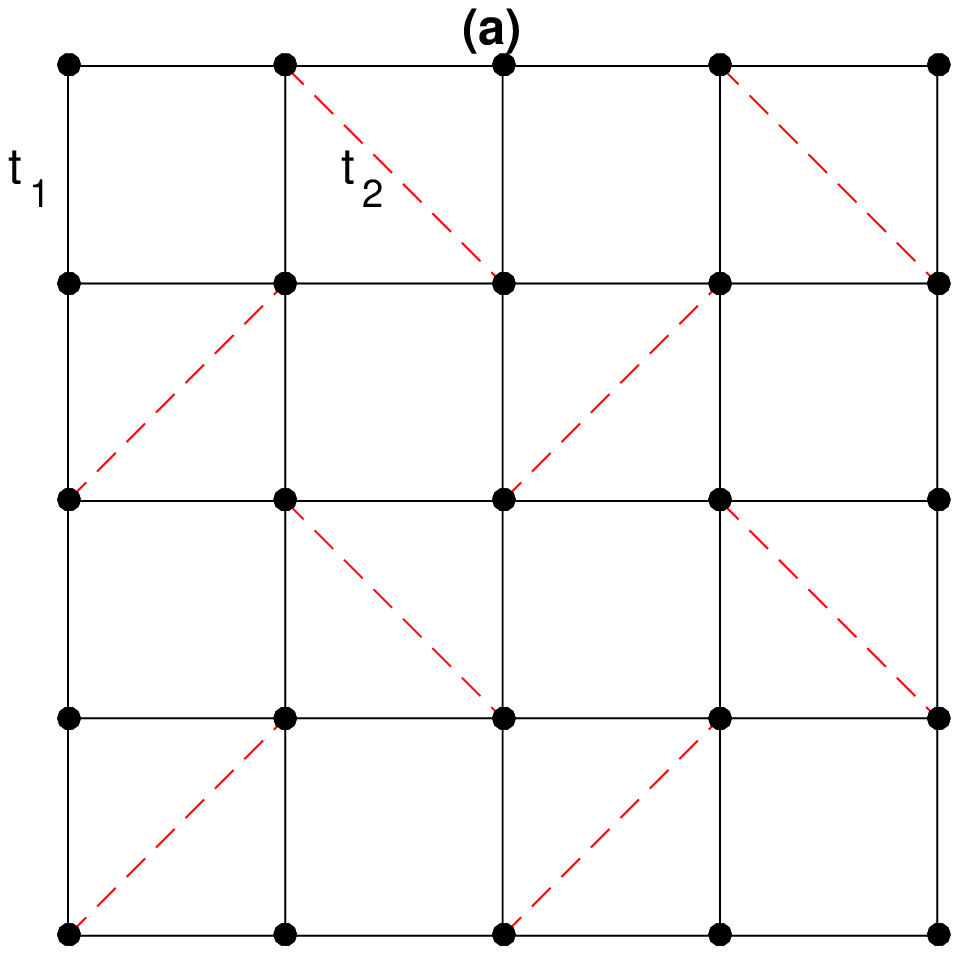}
\includegraphics[width=7.2cm]{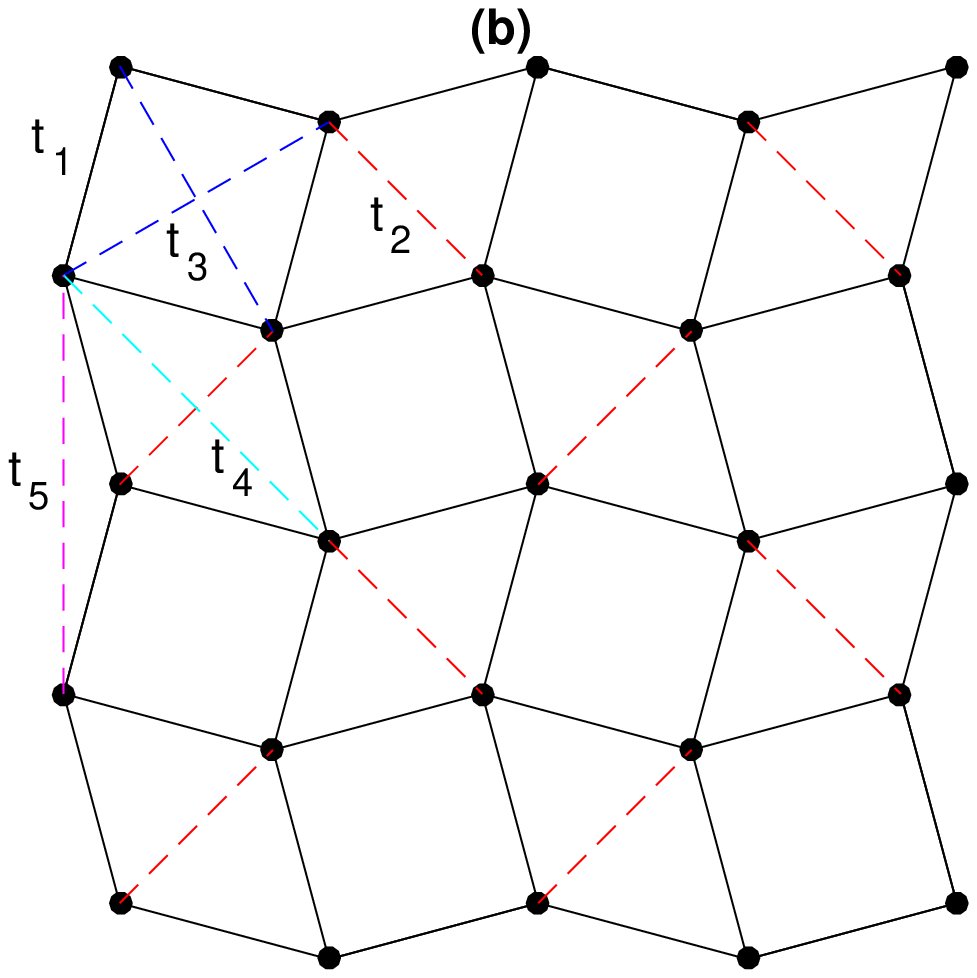}
\end{center}
\caption{\small (a) The original SSL with the first ($J_1$) and second
($J_2$) nearest-neighbor couplings, and (b) the topologically identical structure
realized in the (001) plane of rare-earth tetraborides
with the first ($J_1$), second ($J_2$), third ($J_3$), fourth ($J_4$) and
and fifth ($J_5$) nearest-neighbor couplings.}
\label{fig1}
\end{figure}
The SSL attracted much attention after its experimental realization 
in the $SrCu_2(BO_3)_2$ compound~\cite{Kageyama2, Kodama}. The observation 
of a fascinating sequence of magnetization plateaus ($m/m_s=$1/2, 1/3, 1/4 and 1/8) 
in this material stimulated further theoretical~\cite{Dorier,Corboz} and experimental
studies~\cite{Levy,Takigawa} of the SSL. 
Some time later, many other Shastry-Sutherland  magnets have been discovered. In particular, this 
concerns an entire group of rare-earth metal tetraborides $RB_4$ ($R=La-Lu$). 
These compounds have the tetragonal crystal symmetry $P4/mbm$ with magnetic 
ions $R^{3+}$ located on the SSL in the $ab$ plane carrying a large magnetic moment. 
Moreover, if the crystal field effects are strong enough, then the compounds can be described 
in terms of an effective spin-1/2 Shastry-Sutherland model under strong Ising anisotropy~\cite{Gabani}. 
This is, for example, the case of $TmB_4$~\cite{Gabani,Matas} and $ErB_4$~\cite{Matas, Michimura}, 
where the easy-magnetization axis is normal to Shastry-Sutherland planes. Thus, the study of the Ising limit is 
the first and natural step toward a complete understanding of magnetization processes in these materials.  

The Shastry-Sutherland rare-earth metal tetraborides exhibit similar sequences of fractional magnetization
plateaus as observed in the $SrCu_2(BO_3)_2$ compound. For example, for $ErB_4$ the magnetization 
plateau has been found at $m/m_s=1/2$~\cite{Matas,Michimura}, for $TbB_4$ at $m/m_s=$2/9, 1/3, 4/9, 
1/2  and 7/9~\cite{Yoshii}, for $HoB_4$ at $m/m_s=$1/3, 4/9 and 3/5~\cite{Matas} and for 
$TmB_4$ at $m/m_s=$1/11, 1/9, 1/7 and 1/2~\cite{Gabani}.
As mentioned above, the first attempts to explain the origin of the  fractional magnetization plateaus 
in the metallic Shastry-Sutherland magnets have been made in terms of the Ising model on the
SSL. This model has 
been solved numerically~\cite{Meng,Chang} as well as analytically~\cite{Dublenych1} with a conclusion that 
only the $m/m_s=1/3$ plateau  is stabilized by $J_1$ and $J_2$ interactions. The subsequent analytical 
studies~\cite{Dublenych2} of the model extended by an additional interaction $J_3$ along the diagonals 
of "empty" squares showed that this interaction gives rise to a new magnetization plateau at one-half of 
the saturation magnetization. This result and similar ones obtained~\cite{Suzuki1,Suzuki2} within the 
spin-1/2 XXZ model with the additional ($J_3$) and ($J_4$) interactions pointed to the fact that 
the long-range interactions could play the crucial role in the stabilization of different magnetization 
plateaus with fractional magnetizations. This supposition supports also 
the  recent results of Dublenych~\cite{Dublenych3} and Feng et al.~\cite{Feng}. 
Feng et al. studied the classical Ising model  with the long-range
Ruderman-Kittel-Kasuya-Yosida  (RKKY) interaction on the Archimedean lattice
that is topologically equivalent to SSL one and found that this type of interaction 
leads to the stabilization of new magnetization plateaus  at $m/m_s=1/2$ and $m/m_s=1/4$ 
of the saturated magnetization. Motivated by these works, showing on the importance of
long-range interactions for a correct description of magnetization plateaus
in the Shastry-Sutherland magnets, we have performed in our preceding paper~\cite{pssb2015} 
the systematic study of the influence 
of additional $J_3$ and $J_4$ interactions (as defined in~\cite{Suzuki1}) on the ground states 
and magnetization processes in the extended Ising model. 
For $J_4$=0, we have found that the point $J_3$=0, corresponding to the ordinary
Ising model on the SSL, is the special point of the $J_3-h$ phase diagram with
the only one intermediate plateau at $m/m_s$=1/3, in accordance with previous
numerical~\cite{Meng,Chang} as well as analytical~\cite{Dublenych1} results.
As soon as $J_3$ is nonzero, the new magnetization plateau at $m/m_s$=1/2 is
stabilized against the $m/m_s$=1/3 plateau for both, positive as well as negative
values of $J_3$. At some critical value of $J_3$ interaction, the
$m/m_s$=1/3 plateau completely disappears what accords with
experimental measurements in some rare-earth tetraborides, e.g., 
$ErB_4$~\cite{Matas,Michimura}. However, it should be noted that the  
Ising model with $J_1, J_2$ and $J_3$ interactions is not able to explain
all aspects of magnetization processes in rare-earth tetraborides
and therefore we have extended our model (in accordance with  the theoretical works 
by Suzuki et al.~\cite{Suzuki1,Suzuki2}) by the additional $J_4$ interaction and 
examined  its influence on the formation of magnetization plateaus. It was found 
that the $J_4$ interaction in a combination with the $J_3$ interaction is able to 
generate a number of new magnetization plateaus with the fractional magnetization:
$m/m_s$=1/10, 1/9, 1/6, 1/5, 2/5, 4/9, 7/15 and 5/9
in accordance  with experimental measurements of magnetization curves 
in selected rare-earth tetraborides. However, these results can not be
considered as definite since the $J_4$ interaction is the fourth
nearest-neighbour interaction only for these values of $J_1$ and $J_2$ 
that satisfy the condition $J_1<J_2$ (the case of $SrCu_2(BO_3)_2$
compound), but not for the case $J_1=J_2$ that corresponds to the
real situation in rare-earth tetraborides. In this case the $J_4$
interaction is the fifth nearest neighbor interaction and thus for 
correct description of situation in rare earth tetraborides it 
is necessary to include the true fourth nearest neighbour interaction
as illustrated in Fig.~1b.

Thus our starting Hamiltonian has the form:
\begin{eqnarray}
H&=&J_1\sum_{\langle i,j\rangle_1} S^z_iS^z_j 
+J_2\sum_{\langle i,j\rangle_2} S^z_iS^z_j
+J_3\sum_{\langle i,j\rangle_3} S^z_iS^z_j
\nonumber\\
&+&J_4\sum_{\langle i,j\rangle_4} S^z_iS^z_j+J_5\sum_{\langle i,j\rangle_5} S^z_iS^z_j
- h\sum_i S^z_i\ ,
\label{eq1}
\end{eqnarray}
where $S^z_i=\pm1/2$ denotes the $z$-component of a spin-1/2 degree of freedom on site $i$ of a square lattice 
and $J_1$, $J_2$, $J_3,J_4$ and $J_5$ are the exchange couplings between the
first, second, third, fourth and fifth nearest-neighbour spins on the SSL as indicated 
in Fig.~\ref{fig1}, and $h$ is the magnetic field.

\section{Results and discussion}
To reveal the combined effects of $J_3,J_4$ and $J_5$ interaction on the formation of 
magnetization plateaus in the generalized Ising model on the SSL we have performed the 
exhaustive  studies of the model for a wide range of model parameters. In accordance with 
the previous theoretical as well as experimental works~\cite{Gabani,Chang}, we set $J_1=J_2=1$
and remaining parameters $J_3,J_4$ and $J_5$ are changed from -1 to 1 with a
step 0.05. For each combination of $J_3,J_4$ and $J_5$ the complete
magnetization curve is calculated with a step $\Delta h =0.05$ and 
subsequently the ground-state phase diagram of the model in the $h-J_3$ plane
is presented for representative values of $J_4$ and $J_5$ interactions
to demonstrate effects of long-range interactions on the formation of magnetization
plateaus. We consider both, the positive as well as negative values of 
$J_3,J_4$ and $J_5$, since the recent work by Feng et al.~\cite{Feng} examining 
the influence of  RKKY interaction in these materials showed that the sign of
$J_3,J_4$ and $J_5$ interction does not need to be strictly positive (negative) but 
can change with the value of the Fermi momentum of conduction electrons.
From this point of view our numerical results represent the first systematic study of the influence 
of $J_3,J_4$ and $J_5$ interactions on the formation of magnetization plateaus in the extended Ising 
model. To study the ground-state properties of the model we have used 
the classical Monte Carlo method with the standard Metropolis algorithm~\cite{Newman}. In our 
implementation, the numerical calculations start at finite and sufficiently large temperature 
$T_0$ and the ground states are approached by gradual decreasing of temperature from its initial value. 
To minimize the problem of local minima we perform (for each selected set
of model parameters $J_1,J_2,J_3,J_4,J_5$), $n$ completely independent runs 
(typically twenty), starting from different initial states. For selected 
values of $J_3,J_4$ and $J_5$ interactions we have compared numerical results
obtained within our implementation with ones obtained within a similar
implementation used frequently in the Monte Carlo applications,  
the parallel tempering method~\cite{Earlab}, and we have found that
both methods give practically identical results, what independently confirms 
the reliability of our method. Of course, such a procedure demands in
practice a considerable amount of CPU time, which imposes severe restrictions 
on the size of clusters that can be studied with this method
($L = 24 \times 24$).
Fortunately, we have found that the ground-state energy, as well as  
the width of the magnetization plateaus  depends on $L$ only very weakly 
(for a  wide  range of the model parameters) and thus already such small 
clusters can describe satisfactorily the ground state properties of the model.

Let us first discuss combined effects of $J_3$ and $J_4$ interactions
($J_5=0$). The numerical results obtained for this case are summarized
in Fig.~2 in the form of $h-J_3$ phase diagrams calculated for several
different positive and negative values of $J_4$ on the finite cluster 
of $L=24\times24$ sites. 
\begin{figure}[h!]
\hspace{-1.0cm}
\includegraphics[width=16.0cm]{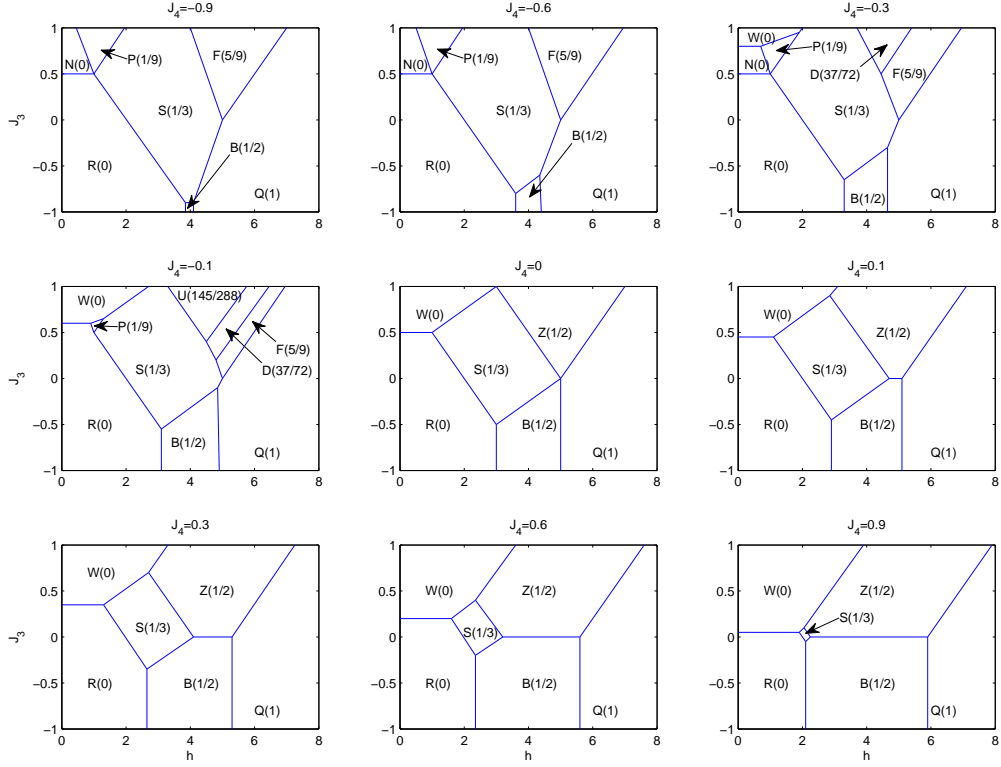}
\caption{\small 
The ground-state phase diagrams of the extended Ising model on the SSL 
in the $J_3-h$ plane calculated for different $J_4$ values on the finite 
cluster of $L=24\times 24$ sites ($J_5=0$). Corresponding ground-state configurations 
are listed in Fig.~\ref{fig3}.}
\label{fig2}
\end{figure}
We first analyse the case $J_4 > 0$, that 
is slightly simpler. Indeed, in this case the increasing $J_4$ interaction
does not generate any new phases (plateaus), but only renormalizes 
the stability regions of phases existing at $J_4=0$. There are six 
different phases (see Fig.~2 and Fig.~3), and namely, two
antiferromagnetic phases with different spin ordering 
(the phase R and W) corresponding to the $m/m_s=0$ plateau, 
three ferrimagnetic phases corresponding to $m/m_s=1/3$ 
(the phase S) and $m/m_s=1/2$ (the phase B and Z) and the ferromagnetic 
phase Q corresponding to $m/m_s=1$. The renormalization effects
are the most pronounced for the $m/m_s=1/3$ phase, that completely 
disappears at some critical value $J^c_4\sim 0.9$. Thus our results 
can yield the answer on the question why the $m/m_s=1/3$ phase is 
present/absent in some rare-earth tetraborides. The combined effects 
of $J_3$ and $J_4$ interactions, discussed above, provide the natural 
explanation of this phenomena.  

The situation in the opposite limit $J_4<0$ is more complex. Numerical 
results performed in this limit showed that phase diagrams  are now much 
richer, where in addition to the above mentioned $m/m_s=1/2$ and $m/m_s=1/3$ 
phases one can find also large domains (the phase P and F) corresponding 
to the $m/m_s$= 1/9 and 5/9 plateau (see Fig.~2) which were also 
observed experimentally in some rare-earth tetraborides (e.g. TmB$_4$). 
One can see that already very  small non-zero values of $J_4<$0 
change  considerably the character of the phase diagram found for $J_4=0$ 
and lead to the stabilization of new phases with $m/m_s$= 1/9 and 5/9
within relatively large domains. With increasing $|J_4|$ these phases are
further stabilized against the $m/m_s=1/2$ phases (B and Z phases), which 
are gradually suppressed and completely disappear from the phase diagram 
at $J^c_4 \sim -0,1$ (the  Z phase) and $J^c_4 \sim -0,9$ (the B phase).      
To reveal the finite-size effects on the stability of these phases
the same calculations (for selected values of $J_4$) have been performed 
also on finite clusters of $L=12\times12$ and $L=36\times36$. It was found
that with exception of two small domains corresponding to the 
$m/m_s$= 145/288 and $m/m_s$= 37/72 phases, the stability regions 
of all remaining phases are independent of $L$ and thus these
diagrams can be satisfactorily extended to the thermodynamic limit
$L \to \infty$.    
\begin{figure}[h!]
\hspace{-1.8cm}
\includegraphics[width=18.0cm]{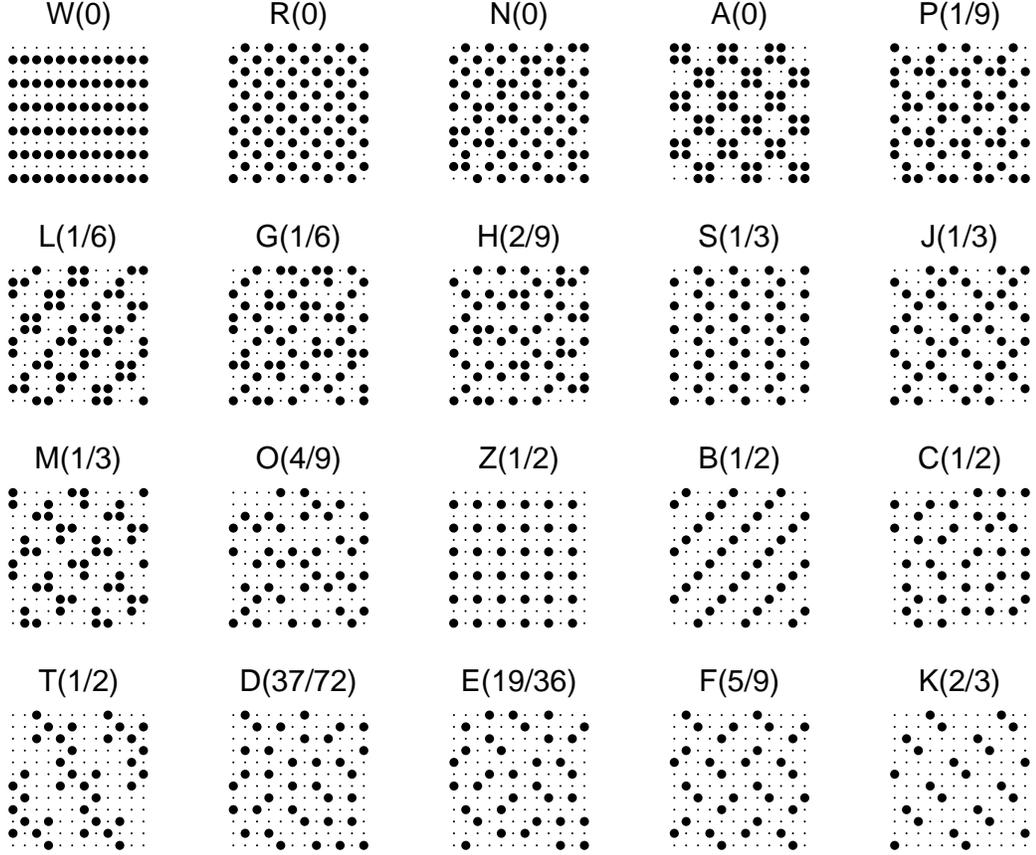}
\caption{\small 
The ground-state configurations detected in the ground-state phase diagrams of the extended Ising model 
on the SS lattice.
}
\label{fig3}
\end{figure}

Now it is interesting to ask what happens with the above described
picture of magnetization processes when the $J_5$ interaction is 
switched on. To  answer this question we have performed the
extensive numerical studies of the model for a wide range of 
model parameters 
$J_3$ (changing from -1 to 1 with a step $\Delta J_3=0.02$), 
$J_4$ (changing from -1 to 1 with a step $\Delta J_4=0.05$) 
and $J_5$ (changing from -0.5 to 0.5 with a step $\Delta J_5=0.1$).
Moreover, taking into account the above mentioned fact, and namely,
that finite-size effects are neglegible for clusters with 
$L \ge 12 \times 12$ all numerical calculations for finite $J_5$
have been performed on the $L=12\times12$ cluster.
Analysing our numerical results we have found that the ground-state 
phase diagrams  are not very sensitive to 
values of $J_4$ interaction and thus in Fig.~4 and Fig.~5 we present 
the phase diagrams only for one representative value 
of $J_4$ interaction ($J_4= -0.6$) and several different values 
of $J_5$ interaction $J_5=0,\pm0.1.\pm0.2,\pm0.3,\pm0.4,\pm0.5$.   
\begin{figure}[h!]
\hspace{-0.5cm}
\includegraphics[width=16.0cm]{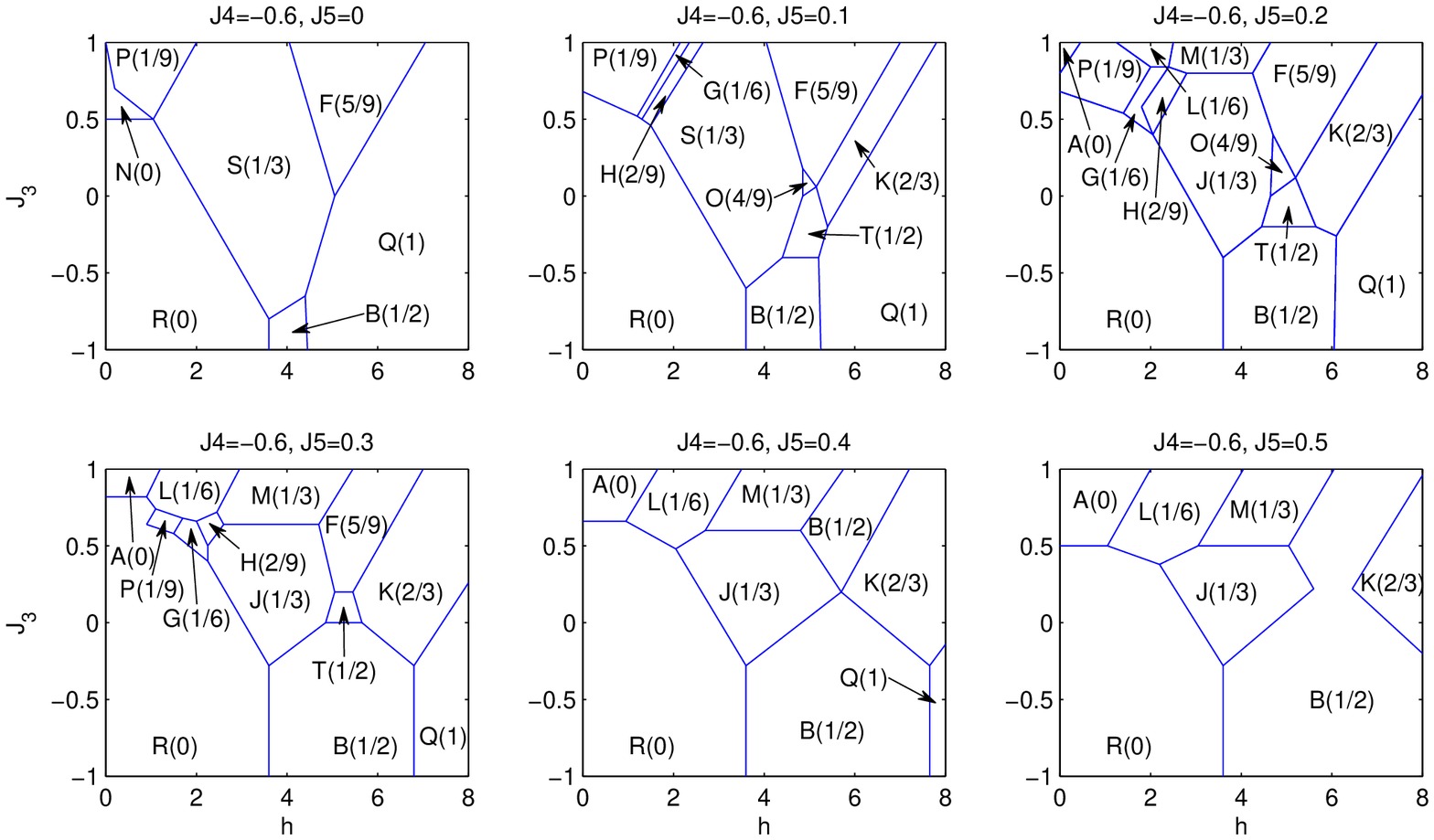}
\caption{\small 
The ground-state phase diagrams of the extended Ising model on the SSL 
in the $J_3-h$ plane calculated for positive $J_5$ values on the finite 
cluster of $L=12\times 12$ sites ($J_4=-0.6$). Corresponding ground-state 
configurations are listed in Fig.~\ref{fig3}.}
\label{fig4}
\end{figure}
One can see that for positive values of $J_5$ interaction the main effect 
of $J_5$ consists in generation of four new magnetization plateaus (phases), 
and namely, the $m/m_s=1/6$, $m/m_s=2/9$, $m/m_s=4/9$ and $m/m_s=2/3$ plateau 
(ground-state spin arrangements corresponding to these phases are displayed in
Fig.~3). 
It should be noted that already very small values of $J_5$ interaction are able to
generate these plateaus and that increasing $J_5$ further stabilizes some
of them. In particular, the $m/m_s=1/6$ and $m/m_s=2/3$ phases are stabilized
against the $m/m_s=1/9$, $m/m_s=2/9$, $m/m_s=4/9$ and $m/m_s=5/9$ phases that 
completely disappear from the phase diagrams for intermediate values of $J_5$
interaction ($J_5\sim 0.4$). Above this value the increasing $J_5$
only slightly renormalizes the width of individual phases but the shape
of the phase diagram, as well as the number of phases remains unchanged. 
Another important result obtained for $J_5>0$ is the existence of  
different spin arrangements corresponding to the "old" 
$m/m_s=0$, $m/m_s=1/3$  and $m/m_s=1/2$  plateaus. Different spin 
configurations corresponding to these plateaus are displayed in Fig.~3 
and phase boundaries between distinct spin arrangements are depicted 
in Fig.~4 by horizontal lines.

The situation in the opposite limit $J_5 < 0$ is shown in Fig.~5
for the same representative values of $J_4$ and $J_5$ interaction. 
\begin{figure}[h!]
\hspace{-0.5cm}
\includegraphics[width=16.0cm]{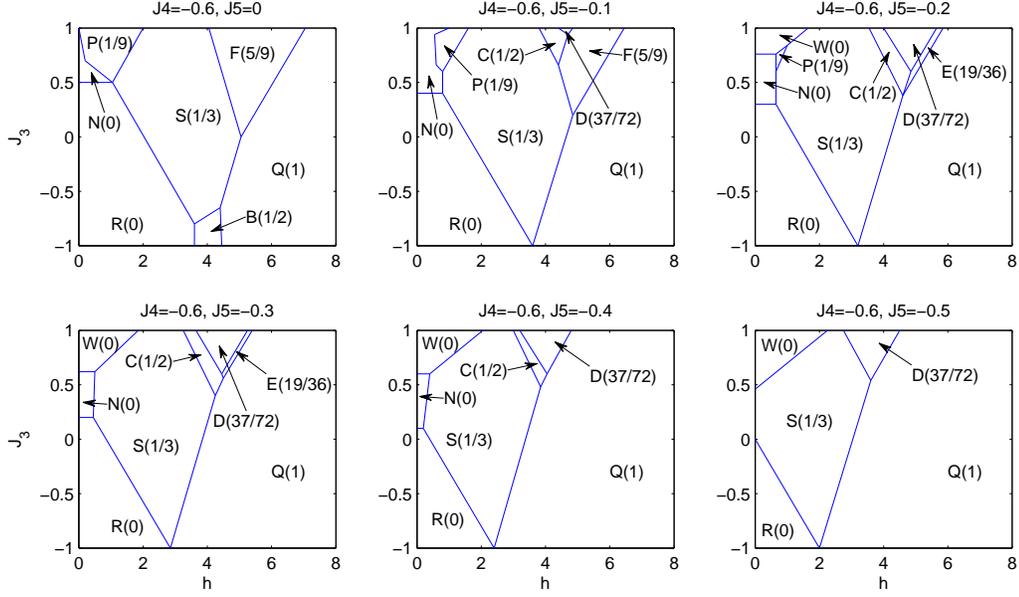}
\caption{\small 
The ground-state phase diagrams of the extended Ising model on the SSL 
in the $J_3-h$ plane calculated for negative $J_5$ values on the finite 
cluster of $L=12\times 12$ sites ($J_4=-0.6$). Corresponding ground-state 
configurations are listed in Fig.~\ref{fig3}.}
\label{fig5}
\end{figure}
Now the main effect of $J_5$ interaction consists in stabilization of 
the $m/m_s=0$ plateau against the $m/m_s=1/9$ plateau and the $m/m_s=1/2$ 
and $m/m_s=37/72$ plateaus against the $m/m_s=5/9$ plateau. Both these 
plateaus are fully suppressed at intermediate values of $J_5$ interaction
$J_5 \sim 0.2$ and above this value the ground-state phase
diagram does not change significantly with $J_5$.

Thus we can summarize that the extended Ising model on the SSL with the 
first, second, third, fourth and fifth nearest-neighbour interactions has a
big potential to explain anomalous magnetic behaviours of a wide group of rare-earth 
tetraborides ($TmB_4$, $TbB_4$, $HoB_4$, $EuB_4$). It provides a reach spectrum of 
magnetic solutions (phases) and corresponding plateaus with the fractional magnetization:
(i) the $m/m_s=0$ plateau for A,N,R and W phases,
(ii) the $m/m_s=1/6$ plateau for L and G phases,
(iii) the $m/m_s=2/9$ plateau for H phase,
(iv) the $m/m_s=1/3$ plateau for S, J and M phases,
(v) the $m/m_s=4/9$ plateau for O phase,
(vi) the $m/m_s=1/3$ plateau for C, B, T  and Z phases,
(vii) the $m/m_s=5/9$ plateau for F phase,
(viii) the $m/m_s=2/3$ plateau for K phase and
(ix) the $m/m_s=1$ plateau for Q phase.  
As was already mentioned in the Introduction some of them have been
observed individually, or in couples within different theoretical
approaches but within our approach we present the whole sequences
of magnetization plateaus and explain under which conditions
the individual plateaus appear or disappear on the magnetization 
curve. Although the model is not able to describe precisely sequences 
of magnetization plateaus as observed in real rare-earth tetraborides, 
it shows that even in such a simple form, may predict the presence 
of many magnetization plateaus that are really observed, and till now 
were not satisfactorily described. Since our numerical results showed
that already very small values of $J_4$ and $J_5$ are able to change
significantly the shape of the magnetization curve found for 
$J_4$ and $J_5$ equal zero, it is natural to ask also on the role 
of additional ($J_6, J_7 \dots $) interactions.
Basically, there are two possible ways how to perform such a generalization.
The first way is to assign the independent interaction constants also
for the additional ($J_6, J_7 \dots $) interactions,
while the second way is to describe the interaction between two spins 
by a simple one-parametric formula, e.g., with exponentially decaying 
interaction amplitudes between $\bf R_i$ and $\bf R_j$ lattice sites, i.e.,
$J_ {ij} = -J_0e^{-\alpha|{\bf R_i}-{\bf R_j}|}$, where $\alpha$ is the
parameter that controls the range of spin interaction and $J_0=e^{\alpha}$.
From the practical point of view, the second method is more suitable 
because it does not expand the model parameter space and has a clearer 
physical meaning, since the  atomic wave functions have also the exponential 
decay with increasing distance. The work in this direction is currently in 
progress.

\vspace{0.5cm}
{\small This work was supported by the Slovak Grant Agency VEGA under Grant
2/0112/18.
Calculations were performed in the Computing Centre of the Slovak Academy
of Sciences using the supercomputing infrastructure acquired in
project ITMS 26230120002 and 26210120002 (Slovak infrastructure for
high-performance computing) supported by the Research and Development
Operational Programme funded by the ERDF.}

\end{document}